\documentclass[conference]{IEEEtran}
\usepackage[letterpaper, portrait, margin=0.75in, top=0.75in, bottom=1.1in]{geometry}
\usepackage[noadjust]{cite}

\usepackage{amsmath,amsfonts}

\usepackage[ruled,vlined]{algorithm2e}

\usepackage{array}
\usepackage{textcomp}
\usepackage{stfloats}
\usepackage{url}
\usepackage{verbatim}
\usepackage{makecell}

\usepackage{xcolor}
\usepackage{amsmath}
\usepackage{amssymb}
\usepackage{mathtools}
\usepackage{lipsum}

\usepackage{subcaption}
\usepackage{cuted}
\begin{document}
\bstctlcite{BSTcontrol}
\title{\huge3D Positioning using a New Diffraction Path Model}

\author{~Gaurav~Duggal, 
~R.~Michael~Buehrer,  ~Harpreet~S.~Dhillon,  and ~Jeffrey~H.~Reed 

\thanks{Authors are with Wireless@VT,  Bradley Department of Electrical and Computer Engineering, Virginia Tech,  Blacksburg,
VA, 24061, USA. Email: \{gduggal, rbuehrer, hdhillon, reedjh\}@vt.edu. }}

% \author{~Gaurav~Duggal \orcid{0000-0002-3765-5479} , 
% ~R.~Michael~Buehrer \orcid{0000-0002-7196-1154},  ~Harpreet~S.~Dhillon \orcid{0000-0003-2829-9449} and ~Jeffrey~H.~Reed \orcid{0000-0003-3494-1901}

% \thanks{G. Duggal, R. M.  Buehrer, H.S. Dhillon and J.H. Reed  are with Wireless@VT,  Bradley Department of Electrical and Computer Engineering, Virginia Tech,  Blacksburg,
% VA, 24061, USA. Email: \{gduggal, rbuehrer, hdhillon, reedjh\}@vt.edu. NIST PSCR PIAP through grant: 70NANB22H070 is gratefully acknowledged.}}

% \vspace{-3mm}
\maketitle
% \vspace{-20mm}

% \author{
% Gaurav Duggal
% \thanks{The authors are with Wireless@VT, Department of ECE, Virginia Tech, Blacksburg, VA. Email: \{gduggal \}@vt.edu. 
% }
% }

% \maketitle

\begin{abstract}
Enhancing 3D and Z-axis positioning accuracy is crucial for effective rescue in indoor emergencies, ensuring safety for emergency responders and at-risk individuals. Additionally, reducing the dependence of a positioning system on fixed infrastructure is crucial, given its vulnerability to power failures and damage during emergencies. Further challenges from a signal propagation perspective include poor indoor signal coverage, multipath effects and the problem of Non-Line-Of-Sight (NLOS) measurement bias. In this study, we utilize the mobility provided by a rapidly deployable Uncrewed Aerial Vehicle (UAV) based wireless network to address these challenges. We recognize diffraction from window edges as a crucial signal propagation mechanism and employ the Geometrical Theory of Diffraction (GTD) to introduce a novel NLOS path length model. Using this path length model, we propose two different techniques to improve the indoor positioning performance for emergency scenarios.
\end{abstract}

% \vspace{-4mm}
\section{Introduction}
In emergency situations, such as fires and mass shooting incidents, the ability to accurately locate both emergency responders and at-risk individuals within a building is of paramount importance. Indoor positioning, particularly for public safety requirements, presents unique challenges, including inadequate indoor signal coverage, a complex signal propagation environment, and the vulnerability of fixed network infrastructure to power loss or damage during emergencies. Furthermore, it is crucial to prioritize enhanced Z-axis positioning performance, as navigating between various floors of a building poses a significant challenge. In the United States, a collaborative solution to tackle this challenge has been established through a public-private partnership between the federal government and AT\&T, known as FirstNet \cite{Firstnetroadmap}. Within FirstNet, a dedicated 20 MHz spectrum in LTE band 14 has been allocated exclusively for public safety communication requirements. 
\par 
This paper focuses on a candidate system for addressing the communication and positioning requirements of public safety networks. The system involves mobile UAVs equipped with position knowledge (and are hence termed {\em anchors}). The mobile anchors establish wireless connections with User Equipment (UE) within a building and have been examined to enhance indoor coverage \cite{duggaletal} by harnessing the mobility provided by the mobile UAVs. Additionally, sensitivity analysis of the parameters controlling positioning performance for this system has been done \cite{dureppagari2023ntn}. The signal propagation environment in this scenario is particularly challenging due to the prevalence of NLOS conditions. Prior research has investigated the exploitation of Multipath Components (MPC) to alleviate the challenges posed by NLOS conditions and enhance positioning accuracy \cite{shen2010fundamental, qi2006analysis}. Subsequently, studies \cite{mendrzik_nlos_mpc, leitinger2015evaluation} have applied Snell's laws to model reflection from planar surfaces, introducing the concept of a `virtual anchor' to enhance positioning performance. In our study, we establish connections between diffraction and positioning, an aspect that, to the best of our knowledge, has received limited attention in the literature. By employing electromagnetic field theory to model diffraction, we formulate a new path length model for NLOS scenarios, utilizing it to devise novel positioning techniques and hence gain additional system level insights.
% \vspace{-2mm}
\section{Problem Formulation and System Model}
\subsection{Window multipath components}
We assume the use of a UAV network with the UAVs acting as base stations referred to as `anchors', deployed outside a building connected to UEs carried by at-risk individuals acting as `nodes' located within the building. Now, we have multiple signal propagation paths between the anchors and the nodes due to the different kinds of interactions with the environment. In general, the signal propagation paths referred to as `multipath components' (MPCs) interact with the environment through three different mechanisms: (a) reflection from flat surfaces, (b) transmission through different materials, or (c) diffraction from edges. 
{\em Bas et al.}  \cite{bas2019outdoor}, conducted a measurement campaign at 28 GHz and observed that for the Outdoor-to-indoor (O2I) scenario, if the incident electric field at the building exterior is at grazing angles of incidence, almost all of the MPCs in the indoor location enter the building through the window. This is due to the fact that brick walls offer high attenuation to the signal.
\par
\begin{figure}[ht]
  % \begin{subfigure}{0.5\textwidth}
  %   \centering
  %   \includegraphics[width=0.8\linewidth]{figs/Building_diffraction_path_EF.PNG}
  %   \caption{Simulation Scenario: UAV-BS at (0,-10,0)}
  % \end{subfigure}%
  \begin{subfigure}{0.48\textwidth}
    \centering
    \includegraphics[width=1\linewidth]{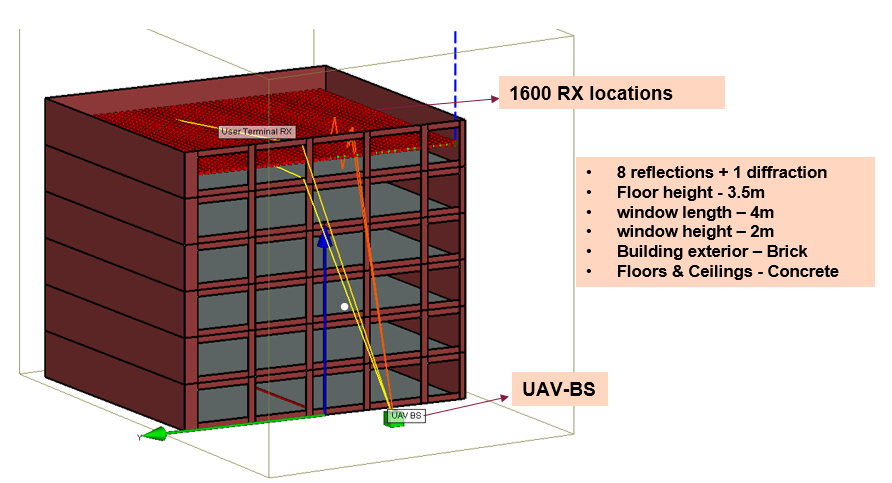}
    % \caption{power ratio of the Diffraction MPC vs elevation angle}
  \end{subfigure}
  \captionsetup{font=small}
  \caption{RayTracing setup of brick building using Remcom's Wireless Insight \cite{wirelessinsight}. }
  \label{fig:raytracing_wireless_insight}
\end{figure}

To further analyze these MPCs due to the window, we created a model of a brick building with glass windows in Remcom’s Wireless Insight RayTracing software
\cite{wirelessinsight} as shown in Fig. \ref{fig:raytracing_wireless_insight}. The transmitter was placed in front of the building on the ground 10m away from the exterior of the building and operated at 28 GHz. We assessed $N_{rx}=1600$ candidate receiver locations on the $5^{th}$ floor of a building with floor dimensions $20m\times 20m$. 
We can represent the path followed by a particular MPC using a string such as `Tx-X-X-...-X-Rx'. Since each MPC begins at the transmitter and terminate at the receiver, all possible strings start with a `Tx' and terminate with an `Rx'. The character `X' in the string are either `R' or `D' representing `Reflection' or `Diffraction'. By reflection we mean specular reflection from a flat surface following Snell's laws whereas diffraction is from edges and follows the diffraction law \cite{namara1990introduction}. The number of characters `X' between the `Tx' and `Rx' represent the number of interactions, and the left to right order of the characters denotes the sequence of interactions. Reflections happen at flat surfaces like ceilings, walls and floors whereas diffraction happens at edges like at the window. There are four main groups of MPCs that are formed by interactions with the windows- MPC-1, MPC-2, MPC-3 and MPC-4. In Table \ref{table_window_MPC}, we see the four possible MPC groups with the corresponding propagation mechanism represented as a string as shown in the second row. 
\par 
For every MPC group, we calculate metrics $P_e$ representing probability of existence of a particular MPC group at a given Rx location, $P_{fap}$ as the probability of the first arriving path being from a particular MPC group. These metrics are shown in the second and third row of the table. Let $N_e$ be the number of Rx locations where a particular MPC group exists and $N_{fap}$ be the number of Rx locations where a particular MPC group is the first arriving path. The first arriving path at an Rx location is defined to be the path which has the shortest propagation delay associated with it. With this we define $P_e$ and $P_{fap}$ as 
\begin{equation}
% %\small
\begin{split}
    P_e = \frac{N_e\times 100}{N_{rx}} ,\;\;\; P_{fap} = \frac{N_{fap}\times 100}{N_{e}}.
\end{split}
\end{equation}
For every MPC group, the third row shows the percentage of Rx locations where the corresponding MPC group exists and the fourth row shows the locations where it is the first arriving path.
\begin{table}[ht]
\captionsetup{font=small}
\caption{MPCs through the windows}
\label{table_window_MPC}
\centering
%\small
\begin{tabular}{|c|c|c|c|c|}
\hline
Group& MPC-1& MPC-2 & MPC-3 & MPC-4   \\
\hline
String& Tx-Rx& Tx-R-Rx & Tx-D-Rx & \makecell{Tx-D-R-...Rx \\ or Tx-R-R-...Rx}  \\
\hline
$P_e$&$1.81\%$&$2.56\%$&$79.375\%$& $100\%$ \\
\hline
$P_{fap}$&$100\%$&$7\%$&$96.69\%$& $16.25\%$ \\
\hline
% $P_{norm}$& \color{red}{To do} & \color{red}{To do}& \color{red}{To do}& \color{red}{To do} \\
% \hline
\end{tabular}
\end{table}
From Table \ref{table_window_MPC} we can conclude that MPC-3 which is formed by a single diffraction from the window edge (Tx-D-Rx), is present for majority of the candidate receiver locations inside the building $P_e=79.375\%$.  Further, it is also the first arriving path for the largest fraction of these receiver locations $P_{fap}=96.69\%$. Motivated by the need to analyze these diffraction paths we develop a simplified building model in the next section.

% \begin{figure*}
%   \begin{subfigure}{0.50\textwidth}
%     \centering
%     \includegraphics[width=1\linewidth]{figs/Law_of_Diffraction.png}
%     \caption{Edge Diffraction using GTD for a half plane}
%    \label{fig_GTD_half_plane}
%   \end{subfigure}%
%   \begin{subfigure}{0.50\textwidth}
%     \centering
%     \includegraphics[width=1\linewidth]{figs/RayTracing_setup.png}
%   \caption{RayTracing setup using Remcom's Wireless Insight\cite{wirelessinsight} with the different MPC groups labelled on a AoA vs path length plot for an RX location on the $5^{th}$ floor of the building near the back. The first arriving path or the shortest paths are the Diffraction paths i.e. MPC group 3}
%   \label{fig_simplified_building_model}
%   \end{subfigure}
%   \caption{}  
%   % \caption{GTD for a half plane and the simplified building model}
%   \label{fig:GTD_geometry}
% \end{figure*}

% \vspace{-2mm}

\subsection{Simplified building model}
\begin{figure*}
  \begin{subfigure}{0.45\textwidth}
    \centering
    \includegraphics[width=1\linewidth]{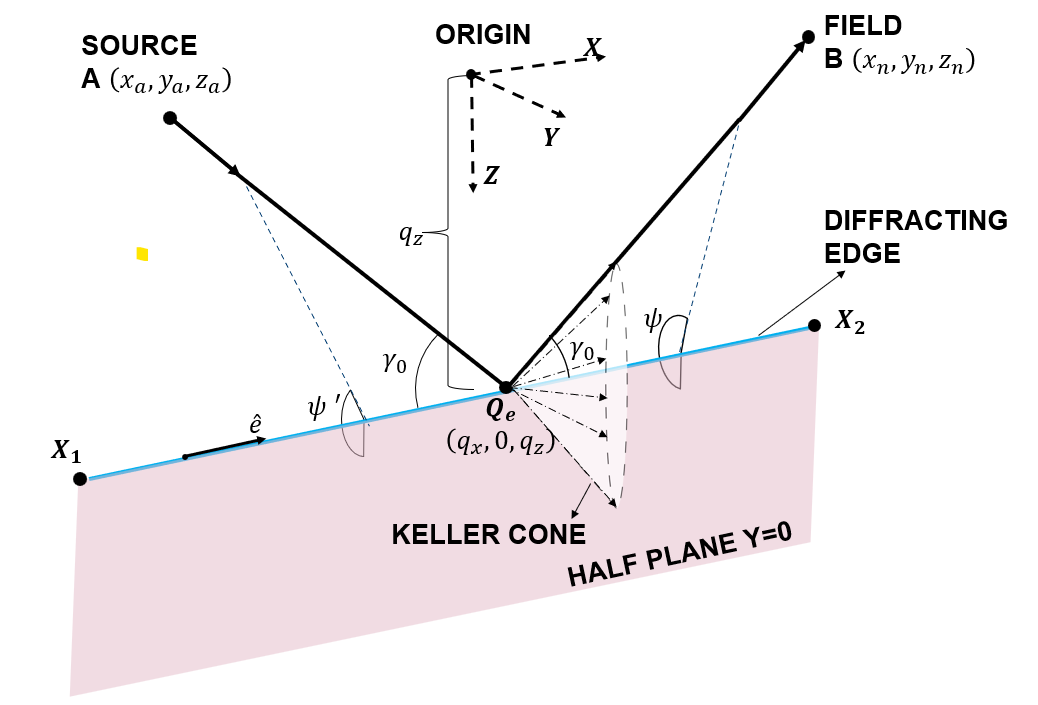}
    \captionsetup{font=small}
    \caption{3D Edge diffraction using GTD for a half plane.}
   \label{fig_GTD_half_plane}
  \end{subfigure}%
  \begin{subfigure}{0.45\textwidth}
    \centering
    \includegraphics[width=1\linewidth]{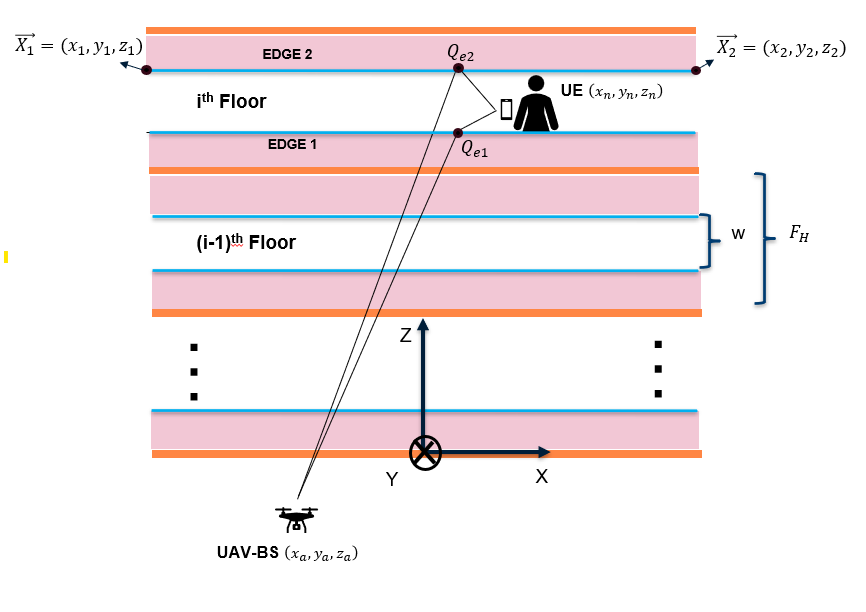}
    \captionsetup{font=small}
  \caption{Front view: Simplified model of the building.}
  \label{fig_simplified_building_model}
  \end{subfigure}
  \captionsetup{font=small}
  \caption{3D Edge diffraction leading to a simplified building model to analyze the diffraction MPCs.}  
  % \caption{GTD for a half plane and the simplified building model}
  \label{fig:GTD_geometry}
\end{figure*}

To develop the simplified building model, we look to model the diffraction from the window edges using an asymptotic technique from electromagnetic field theory called the Geometrical Theory of Diffraction (GTD) \cite{namara1990introduction,balanis2012advanced}. As a canonical example we present diffraction due to an edge. Note in Fig. \ref{fig_GTD_half_plane}, the source A transmits a signal that diffracts from the edge with endpoints $X_1$ and $X_2$ resulting in a cone of diffracting rays called the Keller cone \cite{rahmat2007keller}. Diffraction occurs at the point $\mathbf{Q}_e$, located on the edge and its coordinates are obtained using the law of diffraction \cite{namara1990introduction,balanis2012advanced}. The law states that the angle $\gamma_0$ between the incident ray $\overrightarrow{AQ_e}$ and the edge $\overrightarrow{X_1X_2}$ is preserved after diffraction such that it is equal to the angle between the diffracted ray $\overrightarrow{Q_eB}$ and the edge $\overrightarrow{X_1X_2}$. Mathematically it is expressed as
\begin{equation}
%\small
\label{eq_law_of_diffraction}
    |\overrightarrow{AQ_e} \cdot \overrightarrow{X_1X_2}| = |\overrightarrow{Q_eB} \cdot \overrightarrow{X_1X_2}|.
\end{equation}
Here, $|\overrightarrow{M}\cdot\overrightarrow{N}|$ represents the absolute value of the vector dot product between vectors $\overrightarrow{M}$ and $\overrightarrow{N}$.

% \begin{figure}[ht]
%   % \begin{subfigure}{0.5\textwidth}
%   %   \centering
%   %   \includegraphics[width=0.8\linewidth]{figs/Building_diffraction_path_EF.PNG}
%   %   \caption{Simulation Scenario: UAV-BS at (0,-10,0)}
%   % \end{subfigure}%
%   \begin{subfigure}{0.48\textwidth}
%     \centering
%     \includegraphics[width=0.8\linewidth]{figs/Law_of_Diffraction.png}
%     % \caption{power ratio of the Diffraction MPC vs elevation angle}
%   \end{subfigure}
%   \caption{Edge Diffraction using GTD for a half plane}
%   \label{fig_GTD_half_plane}
% \end{figure}

% \begin{figure}[ht]
%   % \begin{subfigure}{0.5\textwidth}
%   %   \centering
%   %   \includegraphics[width=0.8\linewidth]{figs/Building_diffraction_path_EF.PNG}
%   %   \caption{Simulation Scenario: UAV-BS at (0,-10,0)}
%   % \end{subfigure}%
%   \begin{subfigure}{0.48\textwidth}
%     \centering
%     \includegraphics[width=0.8\linewidth]{figs/Building_model.png}
%     % \caption{power ratio of the Diffraction MPC vs elevation angle}
%   \end{subfigure}
%   \caption{Front view: Simplified model of the building}
%   \label{fig_simplified_building_model}
% \end{figure}

Now, in MPC-3, we have two distinct diffraction paths: one each from the upper and lower horizontal edges of the window frame. Note that for the vertical edges if we apply the law of diffraction \eqref{eq_law_of_diffraction}, we observe that the diffraction point $Q_e$ lies beyond the extremities of the vertical edges of the window. This means there is no edge diffraction from the vertical edge. Instead we need to model diffraction from the corners, i.e., where the horizontal and vertical edges meet \cite{siktaetal}. For the scope of this paper we ignore corner diffraction and therefore can remove the vertical edges. Consistent with these observations we propose a simplified building model as in Fig. \ref{fig_simplified_building_model}, where, each floor of the building consists of two diffracting edges representing the upper and lower horizontal edges of the window.
% \vspace{-5mm}

\section{Path Length Modeling}

\subsection{A new NLOS path length model}
In this section, we develop a new path length model for the diffraction MPCs. For the derivation of the path length we look at the upper edge diffraction MPC. Consider Fig. \ref{fig_simplified_building_model} and let the node be located on the $i^{th}$ floor with the Z-axis coordinate $z_{n,i}$ and (X,Y) coordinate $\boldsymbol{\alpha}_i=[x_n,y_n]^T$. For a building with `$N$' floors, along the vertical axis, $z_{n,i}$ can take `$N$' discrete values since the node is constrained to lie on the lower surface of the `$i^{th}$' building floor. We assume that the vertical coordinate of the node is at the mid point of the `$i^{th}$' building floor. Therefore, $z_{n,i} \in \{(i-1)F_H+\frac{F_H}{2}\, , i\in[1,N]\}$, where $F_H$ is the building floor height in the vertical dimension. Observe that the diffraction point $\mathbf{Q}_e=[q_x,q_y,q_z]^T$ is constrained to lie on the same floor `$i$' as the node. We also assume the window of vertical dimensions `$w$' is located at the vertical mid point of each floor of the building, hence the vertical coordinate of $\mathbf{Q}_e$, $q_z$ is half a window length above the vertical coordinate of the node. Therefore, $q_z=z_{n,i}+\frac{w}{2}$. Let $\mathbf{X}_1 = [x_1,y_1,z_1]^T$ and $\mathbf{X}_2= [x_2,y_2,z_2]^T$ represent the end points of the upper edge of the `$i^{th}$' floor as shown in Fig. \ref{fig_simplified_building_model}. Now, the coordinates of the diffraction point $\mathbf{Q}_e$ can be written as a linear combination of the end points of the upper edge as
\begin{equation}
%\small
    \mathbf{Q}_e = \lambda \mathbf{X}_1 + (1-\lambda) \mathbf{X}_2, \;\;\;\;\;\; 0 \le \lambda \le 1.
\label{eq_diffraction_coordinate}
\end{equation}
The task now is to obtain $\lambda$ to uniquely determine the diffraction point $\mathbf{Q}_e$. With the constraint on the Z-coordinate of the diffraction point and an appropriate choice of the coordinate system, we can assume that the coordinates of the end points of the diffracting edge $y_1=y_2=0$, and $z_1=z_2=(i-1)F_H+\frac{w+F_H}{2}$. From Fig. \ref{fig_GTD_half_plane}, we can obtain the unit vectors corresponding to the incident ray, diffracted ray, and the diffracting edge in Euclidean coordinates. This is then substituted in the law of diffraction in eq. \eqref{eq_law_of_diffraction} to form an equation. On simplification of the equation, we obtain a quadratic expression in the unknown variable $\lambda$. The roots of the quadratic equation represent two possible values of $\lambda$ and are given by:
\begin{equation}
%\small
\begin{split}
\lambda & = \frac{-b \pm \sqrt{b^2-4ac}}{2a}  \\
a &  = (x_1-x_2)^2 \left[(y_n^2-y_{a}^2)+ (z_{n,i}^2-z_a^2)  +2z_1(z_a-z_{n,i})\right] \\
b &  = 2(x_1-x_2) \left[(x_2-x_a)\left((z_1-z_{n,i})^2+y_n^2\right) \right. \\& \left. \;\;\;\;\;\;\;\;\;\;\;\;\;\;\;\;\;\;\;\;\;\;\;\;\;\;\;\;\; -(x_2-x_n)  \left((z_1-z_a)^2+y_a^2\right)\right] \\
c & = (x_2-x_a)^2 \left[(z_1-z_{n,i})^2+y_n^2\right] \\&\;\;\;\;\;\;\;\;\;\;\;\;\;\;\;\;\;\;\;\;\;\;\;\;\;\;\;\;\;   - (x_2-x_n)^2  \left[(z_1-z_a)^2+y_a^2\right].
\label{eq_quadratic_qe}
\end{split}
\end{equation}
Note, only one of the roots satisfies the diffraction law and we drop the other root to obtain a unique solution for the point $\mathbf{Q}_e= [q_x,q_y,q_z]^T$. Here, from eq. \eqref{eq_diffraction_coordinate} we have $q_x=\lambda x_1+(1-\lambda x_2)$, $q_y=y_1=y_2=0$, and $q_z=z_1=z_2$. Also, $q_z$  only takes discrete values corresponding to the `$N$' floors, i.e., one value from $\{(i-1)F_H+\frac{w+F_H}{2}| i \in [1,N]\}$. Assuming we have `$M$' anchors indexed by suffix `$j$' at known locations. The path length $p_{i,j}(\boldsymbol{\alpha}_i)$ corresponding to the upper edge diffraction MPC between the $j^{th}$ anchor and the node located on the $i^{th}$ floor can be expressed as
\begin{equation}
\begin{split}
p_{i,j}(\boldsymbol{\alpha}_i) = \sqrt{(x_{a,j}-q_x)^2+(y_{a,j})^2+(z_{a,j}-q_{z,i})^2} \\ +  \sqrt{(x_n-q_x)^2+(y_n)^2+(0.5w)^2}.
\end{split}
\label{eq_path_length}
\end{equation}
% \vspace{-6mm}
% \vspace{5mm}
\subsection{Range measurements of the diffraction MPCs}
\begin{figure}[ht]
  % \begin{subfigure}{0.5\textwidth}
  %   \centering
  %   \includegraphics[width=0.8\linewidth]{figs/Building_diffraction_path_EF.PNG}
  %   \caption{Simulation Scenario: UAV-BS at (0,-10,0)}
  % \end{subfigure}%
  \begin{subfigure}{0.48\textwidth}
    \centering
    \includegraphics[width=0.8\linewidth]{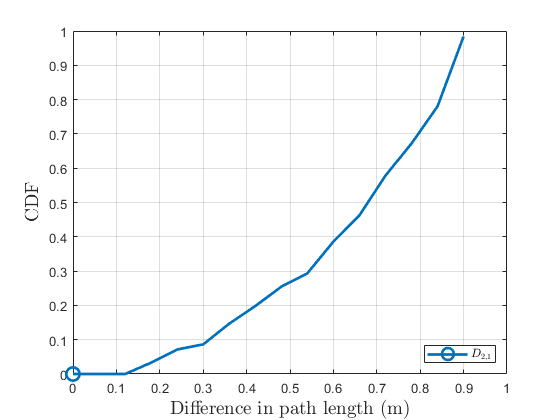}
    % \caption{power ratio of the Diffraction MPC vs elevation angle}
  \end{subfigure}
  \captionsetup{font=small}
  \caption{Difference in path length between the upper edge and lower edge. diffraction paths.}
  \label{fig_path_difference_histogram}
\end{figure}

In this section we seek to establish a link between the range measurement conducted by a Time-of-Flight (TOF) system that relies on estimating the first arriving path in our O2I scenario. We simulated this scenario with four anchors as shown in Fig. \ref{fig_anchor_config} with the simulation parameters in Table \ref{table_simulation_parameters}. Consider a uniformly sampled grid with locations at a distance 0.01m apart across all the floors of the building representing possible node positions. For each node position and for each anchor, we calculated the difference in path length between the upper and lower edge diffraction MPC. In Fig. \ref{fig_path_difference_histogram} we show a histogram (CDF) of the difference in path lengths and observe that the maximum value is 1m. To resolve these paths we would need a bandwidth of at least 300 MHz, which is not realistic in majority of commercial wireless systems. For further theoretical analysis we assume that the TOF based range measurement $r_{l,j}$ between a node present on the $l^{th}$ floor and the $j^{th}$ anchor will correspond to a noisy measurement of the path length corresponding to the upper diffraction path $p_{t,j}(\boldsymbol{\alpha}_t)$. This assumption is relaxed in our evaluation of the performance of the positioning algorithms.

\begin{table}[t!]
\caption{Simulation Parameters}
\label{table_simulation_parameters}
\centering
\begin{tabular}{|c|c|c|c|c|}
\hline
Parameter& Symbol/Value  \\
\hline
 UAVs &$M=4$    \\
\hline
 Floors & $N=7$   \\
\hline
 Floor Height & $F_H=3.5m$  \\
\hline
 Floor Length & $L=20m$   \\
\hline
 Floor Breadth & $B=20m$  \\
\hline
 Window Height & $w=1m$  \\
\hline
\end{tabular}
\end{table}

% \vspace{-2mm}
\section{NLOS 3D Positioning Algorithms}
Positioning in the NLOS setting is particularly challenging due to the presence of NLOS bias in the range measurements. In this section, we present four techniques based on our new path length model and that incorporate varying levels of information about the NLOS paths to improve both the Z-axis position estimate and overall 3D position estimate.  
\begin{figure}[ht]
  % \begin{subfigure}{0.5\textwidth}
  %   \centering
  %   \includegraphics[width=0.8\linewidth]{figs/Building_diffraction_path_EF.PNG}
  %   \caption{Simulation Scenario: UAV-BS at (0,-10,0)}
  % \end{subfigure}%
  \begin{subfigure}{0.40\textwidth}
    \centering
    \includegraphics[width=1\linewidth]{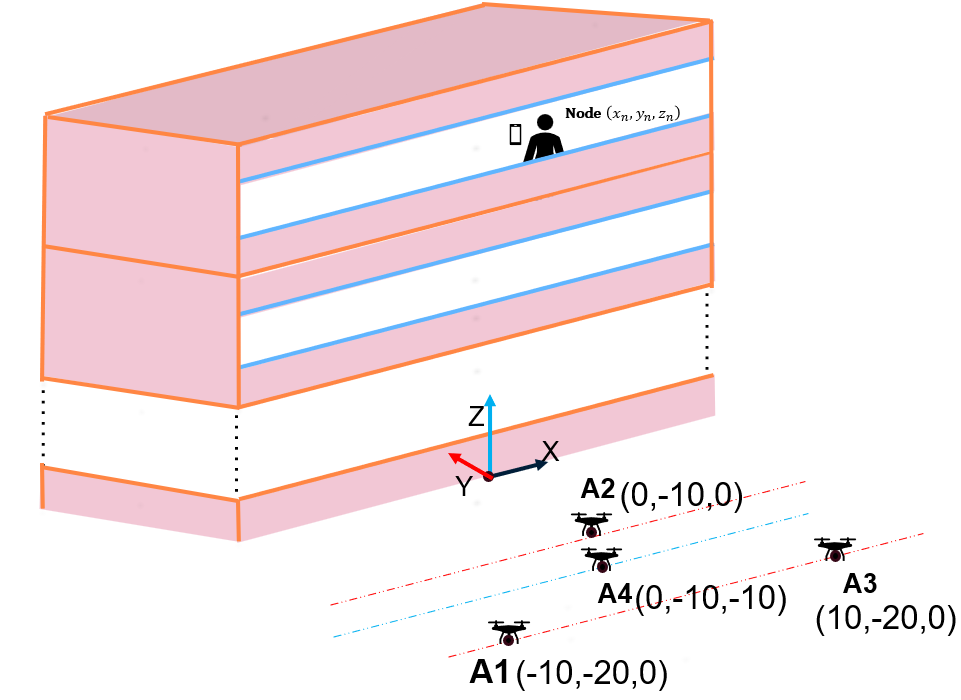}
    % \caption{power ratio of the Diffraction MPC vs elevation angle}
  \end{subfigure}
  \captionsetup{font=small}
  \caption{Anchor configuration for 3D positioning.}
  \label{fig_anchor_config}
\end{figure}
Consider Fig. \ref{fig_anchor_config}, where we have placed $M=4$ UAV anchors on one side of the building. This is done so that we are able to establish a sufficient SNR to nodes located within the building and also to address safety concerns when deploying UAVs around a building. The noisy range measurements  between the four anchors and the node located on an unknown floor `$t$', can be written as a $4 \times 1$ vector of the path length corresponding to the upper edge diffraction MPC as $\mathbf{r}_t = \mathbf{p}_t(\boldsymbol{\alpha}_t) + \mathbf{n}$. The elements of $\mathbf{p}_t(\boldsymbol{\alpha}_t)$ are $p_{t,j}(\boldsymbol{\alpha}_t)$ and are clearly longer than the Euclidean distance separating the anchor and the node. This excess path length is termed as the NLOS bias and adversely affects the positioning estimate. Since the position of the node $\boldsymbol{\alpha}_l$ is not known {\em a priori}, the NLOS bias is often modelled as a random variable following an exponential distribution \cite{lonenlos}. 
% \vspace{-4mm}
\subsection{Stochastic NLOS bias mitigation technique}
\par
Amongst the various NLOS mitigation techniques \cite{venkatesh2007non,vaghefinlossdp,jiacol,jia2010set,guvenc2009survey}, we specifically modify a previously proposed positioning technique termed the {\em Iterative Parallel Projection Algorithm (IPPA)} \cite{jia2010set}, which is a computationally efficient algorithm suitable for deployment on low power mobile platforms such as UAVs and UEs. The algorithm has been improved to mitigate NLOS bias errors by including varying amounts of information about the NLOS paths in \cite{jiacol}. We propose to create a 2D iterative estimator based on IPPA for each floor of the building and after all the floor estimators have converged, we select the the one with the smallest residual that forms the Z-axis estimate. The X,Y position estimates on the floor are obtained from the selected floor estimator.

% \begin{figure}[!ht]
%   \begin{subfigure}{0.45\textwidth}
%     \centering
%     \includegraphics[width=0.8\textwidth]{figs/NLOS_bias_1.PNG}
%   \end{subfigure}%
%   % \begin{subfigure}{0.5\textwidth}
%   %   \centering
%   %   \includegraphics[width=0.8\linewidth]{figs/composite_nlos_bias.png}
%   %   \caption{}
%   % \end{subfigure}
%   \caption{NLOS bias floorwise characterization}
%   \label{fig_nlos_bias} 
% \end{figure}
All of the three primary IPPA algorithms require varying amounts of the information about the NLOS paths. In a) IPPA:NM(ID) - we require identification of the NLOS paths, in b) IPPA:NM(ID,min) - we require additional information, i.e., the lower support of the NLOS bias distribution for the NLOS paths, and finally in c) IPPA:NM(ID,mean) we require the mean of the NLOS bias distribution for the NLOS paths. 
Assuming the node is on an unknown floor `$l$', we obtain the range measurements of the diffraction paths as an $M\times1$ vector $\mathbf{r}_{l}$ from the `M' anchors. Now, the update step for the IPPA algorithms first calculates the contribution $\boldsymbol{\beta}_{i,j}$ for the `$j^{th}$' anchor for the `$i^{th}$' floor estimator as 
\begin{equation}
%\small
\begin{split}
\boldsymbol{\beta}_{i,j} &= \boldsymbol{\alpha}_{i} + ({r}_{l,j}-{r}^b_{i,j}) \frac{\boldsymbol{\alpha}_{i}-\mathbf{X}_j}{\|\boldsymbol{\alpha}_{i}-\mathbf{X}_j\|}, j \in [1,M],\; i \in [1,N]. \\
%r_{i,j} &= \sqrt{(x_{a,j}-x_n)^2+(y_{a,j}-y_n)^2+(z_{a,j}-z_{n})^2} \\
\end{split}
\label{eq_IPPA_steps}
\end{equation}
Here, $\boldsymbol{\alpha}_i$ is the 2D X,Y position estimate of the position for the $i^{th}$ floor which is initialized to an arbitrary location on the same floor, $\mathbf{X}_j$ is the $j^{th}$ anchor position and $\lVert.\rVert$ represents the L2 norm.  We include information about the NLOS biases for the three IPPA:NM algorithms via a correction term $r^b_{i,j}$ in the range measurements $r_{l,j}$. We set $r^b_{i,j}=0$ for IPPA-NM(ID), $r^b_{i,j}=min(b_{i,j})$ for IPPA-NM(ID,min) and $r^b_{i,j}=mean(b_{i,j})$ for IPPA-NM(ID,mean). Here, $b_{i,j}$ is the NLOS bias random variable for the NLOS paths between $j^{th}$ anchor and $i^{th}$ floor and this information can be obtained {\em a priori} using our path model using a technique described in Section \ref{section_nlos_bias_characterization}.
\par Now, for every floor estimator, we iteratively update the $i^{th}$ floor estimate $\boldsymbol{\alpha}_i$ with the contribution $\boldsymbol{\beta}_{i,j}$ from the $j^{th}$ anchor based on the condition that the current position estimate $\boldsymbol{\alpha}_i$ falls within the ranging sphere centered at anchor position $\mathbf{X}_j$ with radius corresponding to its NLOS range measurement $r_{l,j}$. This is expressed as
\begin{equation}
%\small
\begin{split}
\boldsymbol{\alpha}_{i} = &\frac{1}{n(\mathcal{N}_a)}\sum_{j\in \mathcal{N}_a} \boldsymbol{\beta}_{i,j}, \\
\mathcal{N}_a = \{j|j\in\{1,M\}, &(r_{l,j}-r^b_{i,j}) \le \lVert \boldsymbol{\alpha}_i - \mathbf{X}_j \rVert \}. 
\end{split}
\label{eq_IPPA_updates}
\end{equation}
Here, $n(\mathcal{N})$ is the number of elements in set $\mathcal{N}$, and $\boldsymbol{\beta}_{i,j}$ is obtained from eq. \eqref{eq_IPPA_steps}. The residual for each floor estimator $\phi_{i}$ is defined as 
\begin{equation}
%\small
\begin{split}
\phi_{i} &= \frac{1}{M}\sum_{j=1}^{M} \left((r_{l,j}-b_{i,j})\lVert \boldsymbol{\alpha}_i-\mathbf{X}_j\rVert\right). %\;\; j \in [1,M]
\end{split}
\label{eq_IPPA_uav_contribution}
\end{equation}
The iterative loop for each floor estimator is stopped when the absolute difference between consecutive iterations of $\phi_i$ is less than a fixed threshold $\delta$. Once all floor estimators converge, we select the estimate $\boldsymbol{\alpha}_i$ with the smallest residual $\boldsymbol{\phi}_i$, along with its corresponding floor index estimate $i$, to form the 3D position estimate.

% The final algorithm is presented in Algorithm \ref{algo_IPPA}.
% \begin{algorithm}[ht]
% %\small
%     \SetAlgoLined
%     \KwResult{$\hat{\alpha}, l$}
%     \KwData{$N,M, r[1:M], F_H, w, r^b[1:M,1:N], X[1:3,1:M]$}
%     Initialize array $\beta[1\colon 3,1:N]=0$, $\;\; \mathbf{\phi}[1\colon N]=0$ \\
%     \For{i=1 to N}{
%         $z_n \leftarrow (i-1)F_H + \frac{w}{2}$; \\
%         Initialize $x_n$,$y_n$ to arbitrary values on the $i^{th}$ floor \\
%         $\alpha[1:3,i] \leftarrow [x_n,y_n,z_n]^T $ \\
%         \While{True}{
%             $\phi_{old}[i] \leftarrow \phi[i]$ \\
%             $\phi[i] \leftarrow 0$, $N_a=0$ \\
%             \For{j=1 to M}{
%                 $\beta[1:3,j]\leftarrow \alpha[i] + (r[j]-r^b[i,j]) \left(\frac{\alpha[i]-X[1:3,j]}{\lVert\alpha[i]-X[1:3,j]\rVert}\right)$ \\
%                 $\phi[i] \leftarrow \phi[i] + (r[j]-\lVert\alpha[1:3,i]-X[1:3,j]\rVert)^2 $ \\
%                 $r_{i,j} \leftarrow \lVert\alpha[1\colon3,i]-X[1\colon3,j]\rVert$ \\
%                 \If{$(r[j]-r^b[i,j]) < r_{i,j}$}{
%                     $\alpha[1:3,i]  = \alpha[1:3,i] + \beta[1:3,j]$; \\
%                     $N_a = N_a + 1$;
%                 }
%             }
%             $\alpha[1:3,i]  = \frac{\alpha[1:3,i]}{N_a}$, $\phi[i]\leftarrow\frac{\phi[i]}{M}$\\
%             \If{$|\phi[i]-\phi_{old}[i]| \le \delta$}{
%                 break
%             }
%         }
%     }
%     $l = \underset{i}{\mathrm{argmin}}(\phi[i])$ \\
%     $\hat{\alpha} = \alpha[1:3,l]$
%     \caption{NLOS Mitigation Algorithm: IPPA}
%     \label{algo_IPPA}
% \end{algorithm}
\subsection{NLOS bias characterization for the diffraction paths}\label{section_nlos_bias_characterization}
It is challenging to obtain the NLOS bias statistics required for the IPPA algorithms via an empirical measurement campaign since these statistics depend on both the anchor position and the unknown node position. Hence, we present an offline way to obtain the the NLOS bias statistics using our proposed path model. We can characterize the NLOS bias random variable for a given building in two ways - (a) as a floor-wise NLOS bias distribution or, (b) as a single composite NLOS bias distribution. 
The NLOS bias distribution between the $j^{th}$ anchor and the node located on the $i^{th}$ floor can be expressed as a difference of the diffraction path length $p_{i,j}(\boldsymbol{\alpha}_i)$ and the Euclidean distance separating the anchor and node as
\begin{equation}
%\small
\begin{split}
    b_{i,j} = p_{i,j}(\boldsymbol{\alpha}_i)  - \lVert \boldsymbol{\alpha}_i- \mathbf{X}_{j} \rVert.
    %\sqrt{(x_{a,j}-x_n)^2+(y_{a,j}-y_n)^2+(z_{a,j}-z_{n,i})^2}
\end{split}
    \label{eq_nlos_bias}
\end{equation}
To obtain the NLOS bias distribution for a fixed known anchor position we assume the node is located on a fixed floor `$i$', and we uniformly sample $x_n$ and $y_n$ based on the floor dimensions. The histogram of all $b_{i,j}$ thus obtained can be used to approximate the distribution of the NLOS bias between the $i^{th}$ floor for the $j^{th}$ anchor. The required statistics for the different IPPA algorithms can now be derived from this. Similarly, we can characterize the NLOS bias statistics as a composite distribution for a building where we calculate the histogram over all floors. 
% \vspace{-3mm}
\subsection{Non-linear Least Squares (NLS) for the diffraction paths}
In contrast with the previous approach, this approach treats the NLOS bias as a deterministic unknown value and directly estimates the node position based on the noisy range measurements. Again, we propose a 2D X,Y position estimator for each floor of the building. Each floor estimator estimates the coordinates of the diffraction point $\mathbf{Q}_e$ on the upper edge of its floor by minimizing the least squares error on the floor. The final position estimate is obtained by selecting the floor estimator with the smallest residual which is the Z-axis estimate and the 2D X,Y position estimate is provided by the selected floor estimator.
The least squares error minimization problem for the $i^{th}$ floor can be written as
\begin{equation}
% \small
\begin{split}
    \hat{\boldsymbol{\alpha}_i} = \underset{\boldsymbol{\alpha}_i}{\mathrm{argmin}} \sum_{j=1}^M| r_{j,l}-{p}_{i,j}(\boldsymbol{\alpha}_i)| ^2.
    %\sqrt{(x_{a,j}-x_n)^2+(y_{a,j}-y_n)^2+(z_{a,j}-z_{n,i})^2}
\end{split}
    \label{eq_ls_error}
\end{equation}
\vspace{5mm}
 Observe that the path length function ${p}_{i,j}(\boldsymbol{\alpha}_i)$ in eq. \eqref{eq_path_length} is not a linear function. It is a non-linear function of both the unknown 2D X,Y node position $\boldsymbol{\alpha}_l = [x_n,y_n]$, in which the floor index `$l$' itself is unknown, the unknown location of the diffraction point $\mathbf{Q}_e$ on the upper edge of the $l^{th}$ floor, the known anchor positions $\mathbf{X}_j=[x_{a,j},y_{a,j},z_{a,j}]^T$ and the known window dimension $w$. Therefore, we propose to use a non-linear least squares iterative technique called the Gauss-Newton method \cite{zekavat2011handbook,kay1993fundamentals}. The iterative update for each floor with the subscript `$k$' representing the iteration index is given by
\begin{align}
%\small
\boldsymbol{\alpha}_{i,k+1} = \boldsymbol{\alpha}_{i,k} &+ ( \mathbf{H}^T_{i,k} \mathbf{H}_{i,k} )^{-1}\mathbf{H}^T_{i,k} \left(\mathbf{r}_l-\mathbf{p}_i(\boldsymbol{\alpha}_{i,k})\right).
\label{eq_NLS_estimator}
\end{align}
Here, $\mathbf{r}_l$ and $\mathbf{p}_i(\boldsymbol{\alpha}_{i,k})$ are $M\times1$ vectors with elements $r_{l,j}$ and $p_{i,j}(\boldsymbol{\alpha}_{i})$ respectively, and $\mathbf{H}_{i,k}$ is the Jacobian matrix 
\begin{align}
%\small
\mathbf{H}_{i,k} &= \begin{bmatrix} \frac{\partial p_{i,1}(\boldsymbol{\alpha}_{i,k})}{\partial x_n}& \frac{\partial p_{i,1}(\boldsymbol{\alpha}_{i,k})}{\partial y_n} \\
                              \vdots & \vdots  \\  
                             \frac{\partial p_{i,M}(\boldsymbol{\alpha}_{i,k})}{\partial x_n}  & \frac{\partial p_{i,M}(\boldsymbol{\alpha}_{i,k})}{\partial y_n}\end{bmatrix},
\label{eq_Jacobian}
\end{align}
whose elements are given by

\begin{equation}
% \label{eq_jacobian_elements}
%\small
\begin{split}
&\frac{\partial p_{i,j}(\boldsymbol{\alpha}_{i})}{\partial x_n}  = \frac{(q_x-x_a)\frac{\partial q_x}{\partial x_n}}{\sqrt{(q_x-x_{a,j})^2+(y_{a,j})^2+(z_{a,j}-q_{z,i})^2}} \\&\;\;\;\;\;\;\;\;\;\;\;\;\;\;\;\;\;\;\;\;\;\;\;+ \frac{(x_n-q_x)(1-\frac{\partial q_x}{\partial x_n})}{\sqrt{(x_n-q_x)^2+(y_{n})^2+(0.5w)^2}} \\
&\frac{\partial p_{i,j}(\boldsymbol{\alpha}_{i})}{\partial y_n} = \frac{(q_x-x_a)\frac{\partial q_x}{\partial y_n}}{\sqrt{(q_x-x_{a,j})^2+(y_{a,j})^2+(z_{a,j}-q_{z,i})^2}}\\& \;\;\;\;\;\;\;\;\;\;\;\;\;\;\;\;\;\;\;\; + \frac{(q_x-x_n)\frac{\partial q_x}{\partial y_n}+y_n}{\sqrt{(x_n-q_x)^2+(y_{n})^2+(0.5w)^2}}  \\
&\frac{\partial q_x}{\partial x_n} = \frac{(x_1-x_2)}{2a}\left[\frac{\partial a}{\partial x_n}\left(\frac{(b \mp \sqrt{b^2-4ac}}{a}\right)\right.\\&\left.\;\;\;\;\;\;\;\;\;\;\;\;+\left(-\frac{\partial b}{\partial x_n}\pm \frac{b\frac{\partial b}{\partial x_n}-2c\frac{\partial a}{\partial x_n}-2a\frac{\partial c}{\partial x_n}}{\sqrt{b^2-4ac}}\right)\right]   \\\nonumber
\end{split}
\end{equation}

\begin{equation}
\label{eq_jacobian_elements}
%\small
\begin{split}
&\frac{\partial q_x}{\partial y_n} = \frac{(x_1-x_2)}{2a}\left[\frac{\partial a}{\partial y_n}\left(\frac{(b \mp \sqrt{b^2-4ac}}{a}\right)\right.\\&\left.\;\;\;\;\;\;\;\;\;\;\;\;+\left(-\frac{\partial b}{\partial y_n}\pm \frac{b\frac{\partial b}{\partial y_n}-2c\frac{\partial a}{\partial y_n}-2a\frac{\partial c}{\partial y_n}}{\sqrt{b^2-4ac}}\right)\right] \\
&\frac{\partial a}{\partial x_n}  = 0, \;\;
\frac{\partial a}{\partial y_n}  = 2y_n(x_1-x_2)^2  \\ 
&\frac{\partial b}{\partial x_n}  = 2(x_1-x_2)\left((z_1-z_a)^2+y_a^2\right) \\ &\frac{\partial b}{\partial y_n}  = 4y_n(x_1-x_2)(x_2-x_a) \\ 
&\frac{\partial c}{\partial x_n}  = 2(x_2-x_n)\left((z_1-z_a)^2+y_a^2\right)\\ &\frac{\partial c}{\partial y_n}  = 2y_n(x_2-x_a)^2. \\ 
\end{split}
\end{equation}

% Defining the residual for each floor estimator as the L2 norm $\phi_i = \lVert\mathbf{r}_l - \mathbf{p}_i(\boldsymbol{\alpha}_{i}) \rVert$ and after running each estimator for $K$ iterations, the estimate of the floor index $\hat{l}$ is obtained by
% \begin{equation}
% %\small
% \hat{l}= \underset{i}{\mathrm{argmin}} [\phi_1,\cdots,\phi_i,\cdots,\phi_N].
% \end{equation}

% \vspace{-5mm}
\section{Positioning Performance}
In our simulation, we considered 100,000 trials where for each trial we uniformly sampled a 2D position on an arbitrary floor of an `$N$' floored building to obtain range measurements for `$M$' UAV anchors. The other simulation parameters are given in Table \ref{table_simulation_parameters}. Our baseline positioning performance is compared with the traditional Linear Least Squares (LLS)  \cite{zekavat2011handbook}. In our simulation, for each trial and for each range measurement we select the lower or the upper edge with 50\% probability. This includes the effect of noise in the range measurements $\mathbf{r}_l$. 
\par
\begin{figure*}[t!]
     \centering
     \begin{subfigure}[b]{0.24\textwidth}
         \centering
         \includegraphics[clip, trim=1.8cm 0cm 2.25cm 0.7cm, width=\textwidth]{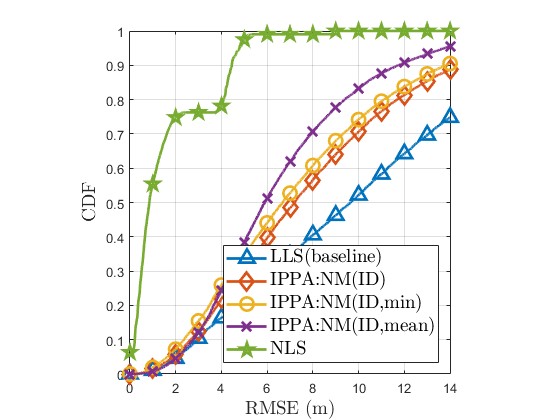}
         \captionsetup{font=small}
         \caption{\footnotesize Floorwise NLOS, 3D-RMSE.}
         \label{fig:rmse_floorwise_3drmse}
     \end{subfigure}
     \hfill
     \begin{subfigure}[b]{0.24\textwidth}
         \centering
        \includegraphics[clip, trim=1.8cm 0cm 2.25cm 0.7cm, width=\textwidth]{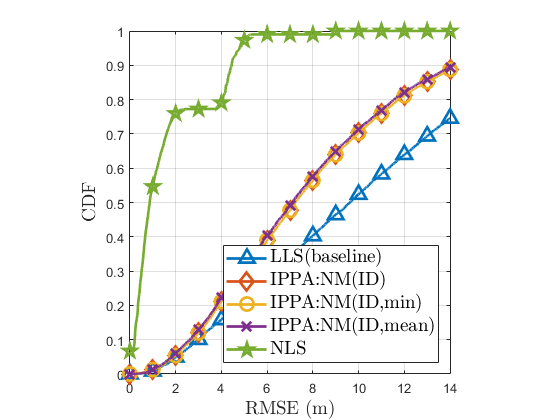}
        \captionsetup{font=small}
        \caption{\footnotesize Composite NLOS, 3D-RMSE.}
        \label{fig:rmse_composite_3drmse}
     \end{subfigure}
     \hfill
     \centering
     \begin{subfigure}[b]{0.24\textwidth}
         \centering
         \includegraphics[clip, trim=1.8cm 0cm 2.25cm 0.7cm, width=\textwidth]{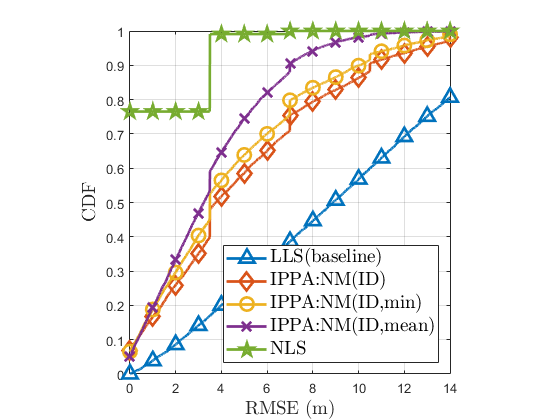}
         \captionsetup{font=small}
         \caption{\footnotesize Floorwise NLOS, Z-RMSE.}
         \label{fig:rmse_floorwise_zrmse}
     \end{subfigure}
     \hfill
     \begin{subfigure}[b]{0.24\textwidth}
         \centering
        \includegraphics[clip, trim=1.8cm 0cm 2.25cm 0.7cm, width=\textwidth]{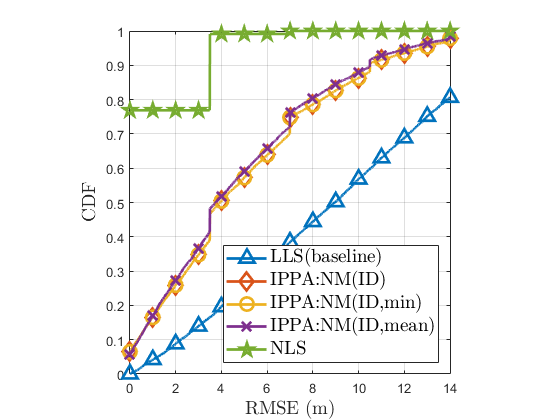}
        \captionsetup{font=small}
        \caption{\footnotesize Composite NLOS, Z-RMSE.}
        \label{fig:rmse_composite_zrmse}
     \end{subfigure}
     \captionsetup{font=small}
     \caption{Positioning Performance: NLS offers the best 3D and Z-axis positioning performance followed by IPPA:NM(ID,mean), IPPA:NM(ID,min) and IPPA:NM(ID) respectively, all of which are better than the baseline LLS approach. Characterization of NLOS bias on a floorwise basis is better than a composite characterization.}
     \label{fig:rmse_algorithms}
\end{figure*}

The performance is evaluated by calculating the Root Mean Squared Error (RMSE) between the ground truth position and the estimated position. The CDF of the 3D-RMSE estimation errors and Z-axis estimation errors are compared in Fig. \ref{fig:rmse_algorithms}. Observe that NLS offers the best 3D-RMSE and Z-RMSE performance but requires {\em a priori} knowledge of the coordinates of the diffracting edges for each floor. The next best performing algorithms are the IPPA algorithms in descending order - (a) IPPA:NM(ID,mean), (b) IPPA:NM(ID,min) and (c) IPPA:NM(ID). This is consistent with the level of {\em a priori} knowledge about the NLOS bias. Finally, on comparing Fig. \ref{fig:rmse_floorwise_3drmse} and Fig. \ref{fig:rmse_floorwise_zrmse} with Fig. \ref{fig:rmse_composite_3drmse} and Fig. \ref{fig:rmse_composite_zrmse}, observe that it is better to characterize the NLOS bias on a floorwise basis to improve both 3D as well as Z-axis positioning performance. An important point to note is despite the poor anchor geometry with all anchors located on one side of the building, the NLS technique offers $<2m$ 3D-RMSE performance at 80\% of the indoor locations. The baseline traditional LLS technique performs poorly due to the presence of NLOS bias.
% \vspace{50mm}
% \vspace{-2mm}
\section{Conclusion and Future Work}
Motivated by the significance of diffraction in the O2I scenario, particularly from the context of positioning, we have developed a new model for NLOS path length, marking the first attempt to incorporate the Diffraction propagation mechanism into TOF-based positioning literature. Subsequently, employing this path length model, we introduced two positioning algorithms designed to enhance both 3D and Z-axis indoor positioning performance by including varying amounts of {\em a priori} information about the NLOS range measurements. Demonstrating the efficacy of the developed algorithms, we illustrated that characterizing the NLOS bias distribution on a floorwise basis for a building significantly improves both 3D and Z-axis positioning as compared to treating it as a single composite distribution. In future investigations, we aim to explore the application of the new path length model to other scenarios involving diffraction, such as outdoor urban environments. Additionally, we intend to assess the impact of anchor geometry on positioning performance.    

\section{Acknowledgements}
The authors would like to gratefully acknowledge NIST PSCR PIAP through grant: 70NANB22H070, NIJ graduate research fellowship through grant: 15PNIJ-23-GG-01949-RESS and NSF through grants CNS-1923807 and CNS-2107276 and Ayushi Puri for help with the diagrams. 
% \begin{equation}
% %\small
% \begin{split}

% \end{split}
% \end{equation}

% \begin{equation}
% %\small
% \begin{split}
% \frac{\partial q_x}{\partial y_n} &= \frac{(x_1-x_2)}{2a}\left[\frac{\partial a}{\partial y_n}\left(\frac{(b \mp \sqrt{b^2-4ac}}{a}\right)\right.\\&\left.\;\;\;\;\;\;\;\;\;\;\;\;+\left(-\frac{\partial b}{\partial y_n}\pm \frac{b\frac{\partial b}{\partial y_n}-2c\frac{\partial a}{\partial y_n}-2a\frac{\partial c}{\partial y_n}}{\sqrt{b^2-4ac}}\right)\right] \\
% \frac{\partial a}{\partial x_n} & = 0, \;\;
% \frac{\partial a}{\partial y_n}  = 2y_n(x_1-x_2)^2  \\ 
% \frac{\partial b}{\partial x_n} & = 2(x_1-x_2)\left((z_1-z_a)^2+y_a^2\right), \\
% \frac{\partial b}{\partial y_n}  &= 4y_n(x_1-x_2)(x_2-x_a)  \\
% \frac{\partial c}{\partial x_n} & = 2(x_2-x_n)\left((z_1-z_a)^2+y_a^2\right)\\
% \frac{\partial c}{\partial y_n}  &= 2y_n(x_2-x_a)^2.  
% \end{split}
% \end{equation}
% \vspace{-2mm}

\bibliography{refs}
\bibliographystyle{IEEEtran}
% \printbibliography

\end{document}